\documentclass[12pt, preprint]{aastex}

\shorttitle{UY Aur at N-band}
\shortauthors{Skemer et al.}

\begin{document}

\title{ISM Dust Grains and N-band Spectral Variability in the Spatially Resolved Subarcsecond Binary UY Aur\footnote{The observations reported here were partially obtained at the MMT Observatory, a facility operated jointly by the Smithsonian Institution and the University of Arizona.} \footnote{The observations reported here were partially obtained at the Gemini Observatory, which is operated by the Association of Universities for Research in Astronomy, Inc., under a cooperative agreement with the NSF on behalf of the Gemini partnership: the National Science Foundation (United States), the Science and Technology Facilities Council (United Kingdom), the National Research Council (Canada), CONICYT (Chile), the Australian Research Council (Australia), CNPq (Brazil), and CONICET (Argentina)} \footnote{The observations reported here were partially obtained at the Infrared Telescope Facility, which is operated by the University of Hawaii under Cooperative Agreement no. NCC 5-538 with the National Aeronautics and Space Administration, Science Mission Directorate, Planetary Astronomy Program.}}

\author{Andrew J. Skemer, Laird M. Close, Philip M. Hinz, William F. Hoffmann}
\affil{Steward Observatory, Department of Astronomy, University of Arizona, Tucson, AZ 85721}

\author{Thomas P. Greene}
\affil{NASA Ames Research Center, Moffett Field, CA 94035}

\author{Jared R. Males}
\affil{Steward Observatory, Department of Astronomy, University of Arizona, Tucson, AZ 85721}

\author{Tracy L. Beck}
\affil{Space Telescope Science Institute, Baltimore, MD 21218}

\begin{abstract}
The 10$\micron$ silicate feature is an essential diagnostic of dust-grain growth and planet formation in young circumstellar disks.  The Spitzer Space Telescope has revolutionized the study of this feature, but due to its small (85cm) aperture, it cannot spatially resolve small/medium separation binaries ($\lesssim$3"; $\lesssim420$ AU) at the distances of the nearest star-forming regions ($\sim$140 pc).  Large, 6-10m ground-based telescopes with mid-infrared instruments can resolve these systems.  

In this paper, we spatially resolve the 0.88" binary, UY Aur, with MMTAO/BLINC-MIRAC4 mid-infrared spectroscopy.  We then compare our spectra to Spitzer/IRS (unresolved) spectroscopy, and resolved images from IRTF/MIRAC2, Keck/OSCIR and Gemini/Michelle, which were taken over the past decade.  We find that UY Aur A has extremely pristine, ISM-like grains and that UY Aur B has an unusually shaped silicate feature, which is probably the result of blended emission and absorption from foreground extinction in its disk.  We also find evidence for variability in both UY Aur A and UY Aur B by comparing synthetic photometry from our spectra with resolved imaging from previous epochs.  The photometric variability of UY Aur A could be an indication that the silicate emission itself is variable, as was recently found in EX Lupi.  Otherwise, the thermal continuum is variable, and either the ISM-like dust has never evolved, or it is being replenished, perhaps by UY Aur's circumbinary disk.
\end{abstract}

\section{Introduction}
Grain growth, which is one of the initial steps to planet formation, can be studied by observing the 10$\micron$ silicate features of young stars.  By modeling the shape and strength of the feature, it is often possible to determine the different dust grain constituents (amorphous grain-size, crystallinity, and in some cases mineralogy) of a young circumstellar disk \citep{2001A&A...375..950B,2005A&A...437..189V,2006ApJ...639..275K,Watson, Sargent}.  Numerous authors have also used simple ``size/shape" silicate feature metrics to determine the characteristic grain-size and crystallinity of dust \citep{2003A&A...400L..21V,2003A&A...412L..43P,2005Sci...310..834A,2006ApJ...639..275K,Ilaria}.  Based on this morphological classification, a logical evolutionary sequence involves small, amorphous, ISM-like grains combining and annealing into large \citep[or more porous---see][]{2008A&A...483L...9V} and crystalline grains on the way to forming planetesimals and eventually planets.  For a complete review, see \citet{2007prpl.conf..767N}.

The Spitzer Space Telescope and its IRS spectrograph have observed numerous young stellar objects, including the \citet{Furlan} Taurus-Aurigae survey of 111 Class II and III objects.  However, a large fraction of these results are contaminated by unresolved binaries due to Spitzer's small (85cm) aperture.  \citet{Furlan} have noted that along with a large number of ``normal" silicate spectra, they have 15 outlier spectra, some of which may be unresolved blends.  Known binaries exist in both the normal subsample and the outlier subsample.  Complementing the space-based results is a spatially-resolved, N-band ($\sim$8$\micron$-13.5$\micron$) imaging survey of binaries by \citet{McCabe}.  However, there is still a need for resolved spectroscopy of these binaries.

In this paper, we present resolved, broad N-band spectroscopy of the 0.88" binary UY Aur using the MMT deformable secondary adaptive optics system (MMTAO), and the mid-infrared camera, BLINC-MIRAC4.  We also present unpublished resolved imaging data from two earlier epochs using IRTF/MIRAC2 and Gemini/Michelle and compare our results to published Keck/OSCIR photometry \citep{McCabe} and Spitzer/IRS blended spectroscopy \citep{Furlan,Watson,Ilaria}.  Collectively, these observations span more than a decade.

UY Aur is a particularly interesting T Tauri system in the Taurus-Aurigae star forming region that was originally discovered to be a binary in the optical \citep{1944PASP...56..123J}, before one of its components, UY Aur B, faded to more than 5 magnitudes fainter than the primary at R-band \citep{1995ApJ...444L..93H}.  The system was observed to be a binary in the infrared by \citet{1993AJ....106.2005G} and \citet{1993A&A...278..129L}, and subsequent H-band and K-band imaging revealed that the system is highly variable at those wavelengths too \citep{1995ApJS..101..117K,1998ApJ...499..883C,2001ApJ...556..265W,2003AJ....126.2009B,2007AJ....134..880H}.  The source of the variability, while assumed to be mostly variable accretion/extinction around UY Aur B may come from UY Aur A as well \citep{2003AJ....126.2009B}.  However, spatially resolved, absolute calibration data from \citet{1998ApJ...499..883C} and \citet{2007AJ....134..880H} show that UY Aur A's H-band flux stayed nearly constant ($\Delta H=$0.4 magnitudes) for three data points between 1996 and 2005, while UY Aur B varied by a maximum of $\Delta H=$1.3 magnitudes.

\citet{1996A&A...309..493D} and \citet{1998A&A...332..867D} resolved a Keplerian circumbinary disk around UY Aur A+B using $^{13}$CO millimeter interferometry.  The disk was subsequently resolved in scattered light by \citet{1998ApJ...499..883C} and polarized scattered light by \citet{2000ApJ...540..422P}.  Using scattered-light multi-wavelength data, \citet{1998ApJ...499..883C} showed that the dust composition of the circumbinary disk was dominated by small (0.03$\micron$-0.6$\micron$) grains and postulated, based on their images, that  spiral-like streamers from the circumbinary disk were replenishing small grains in the circumstellar disks of the binary components.  The circumstellar disks are thought to be truncated to $\lesssim$1/2 the separation of the binary ($\lesssim$60 AU assuming a face-on orbit) based on dynamical simulations \citep{1994ApJ...421..651A}.  The system also has a bipolar outflow, although which star is its source is unknown \citep{1997A&AS..126..437H}.

\citet{2003ApJ...583..334H} used the Space Telescope Imaging Spectrograph (STIS) to acquire optical spectra of both binary components.  Their results show that UY Aur A is an M0 classical T Tauri star with a mass of 0.6 $M_\sun$ and a fairly high accretion rate of $10^{-7.64} M_\sun/yr$.  UY Aur B was characterized as an M2.5 classical T Tauri star with a mass of 0.34 $M_\sun$ and a similarly high accretion rate of $10^{-7.70} M_\sun/yr$  Extinction measurements showed that UY Aur A was extincted by $A_{V}$=0.55 and UY Aur B was extincted by $A_{V}$=2.65.  This large disparity in the extinction of Class II binaries is uncommon since both stars lie along a common line of sight in the dark cloud.  Possible causes of this phenomenon are extinction of UY Aur B from a foreground object (such as the disk of UY Aur A), or self-extinction from the disk (caused, for example, by UY Aur B's disk potentially being viewed from close to edge-on).

In summary, UY Aur is a binary classical T Tauri system with two truncated circumstellar disks.  UY Aur B is an infrared companion that has exhibited variability in the optical and near-IR.  UY Aur A may be variable in the near-IR as well.  The system is surrounded by a circumbinary disk, which may contain spiral-like streamers to the circumstellar disks.

Of the 111 young stars in \citet{Furlan}'s sample, UY Aur and LkCa 15 have spectra that are the best examples of pristine, ISM-like silicate features \citep{Watson}, which could be the result of a lack of dust-grain evolution or variable and/or replenishing mechanisms.  Given that these Spitzer measurements are of a blended spectrum, and UY Aur B is heavily extincted at optical and near-IR wavelengths, it is surprising that \citet{Watson} would measure such a pristine (de-extincted) ISM-like feature.  Our resolved spectroscopy will help determine the nature and differences of the dust components in UY Aur A and B.

\section{Observations and Reductions}

\subsection{MMTAO/BLINC-MIRAC4 AO Spectra}
On January 14, 2009 UT, we observed UY Aur with MMTAO/BLINC-MIRAC4 using MIRAC4's newly commissioned spectroscopic mode.  MMTAO is the 6.5 meter MMT's deformable secondary adaptive optics system \citep[e.g.][]{2000PASP..112..264L, 2003SPIE.5169...17W, 2004SPIE.5490...23B}.  BLINC-MIRAC4 (Bracewell Infrared Nulling Cryostat; Mid-IR Array Camera, Gen. 4) is a combined AO-optimized mid-IR imager (MIRAC4) and nulling interferometer (BLINC), which for these observations, is used in its ``imaging" mode.  MIRAC4 is functionally equivalent to previous versions of MIRAC \citep[e.g.][]{1998SPIE.3354..647H}, with its new features documented in Hinz et al (in prep).  BLINC is descibed in \citet{2000SPIE.4006..349H}.  Using AO, BLINC-MIRAC4 can achieve Strehls of up to $\sim$98\% at 10$\micron$ \citep{2003ApJ...598L..35C}.

MIRAC4 was first used in its spectroscopic mode for grism trace-centroiding, without a flux calibration \citep{Skemer09}.  This paper presents the first flux-calibrated spectra with MIRAC4, and thus a full description of our observing strategies and reductions will follow.

MIRAC4 operates at a variety of platescales ranging from 0.055"/pixel to 0.110"/pixel.  In order to observe the full N-band range (7.6$\micron$-13.3$\micron$), we operate close to the coarsest platescale (0.105"/pixel).  For all the spectra in this paper, we use an oversized (1") slit, to negate object alignment effects.  In this configuration, MIRAC4 has a spectroscopic range of 7.6$\micron$-13.3$\micron$, which is limited by an N-band filter to block higher order grism modes.  Our wavelength calibration was done in a lab the same week as the observations and resulted in a measured dispersion of 27 nm/pixel.  No telluric wavelength calibration is done at this time because of the lack of available lines in the N-band atmospheric window.  Because we use an oversized slit, our spectral resolution is limited by diffraction.  At the center of N-band (10.5$\micron$), the diffraction limit of the 6.5 meter MMT is $\sim\lambda/D=$0.32".  This corresponds to a spectral resolution of $\sim$125.

Conditions on January 14, 2009 UT were photometric at the beginning of the night and we observed the mid-IR standard, $\epsilon$ Tau, at two different airmasses in broad-N spectra, followed by the science target, UY Aur.  Ideally, the UY Aur observations would have been followed by another observation of $\epsilon$ Tau.  However, the wind shifted and strengthened after our UY Aur observation, precluding a third observation of the standard.  All of our observations used adaptive optics with a loop-speed of 550 Hz.  Details of our observations are presented in Table \ref{All Observations}.

Each object was aligned in the slit for two nod beams separated by 5" (we chop perpendicular to the slit with an 8" throw).  Because we were observing the binary UY Aur, we rotated the camera with respect to the sky to manually align the binary with the slit.  The telescope rotator was turned on to maintain the position angle.

\begin{figure}
 \includegraphics[angle=90,width=\columnwidth]{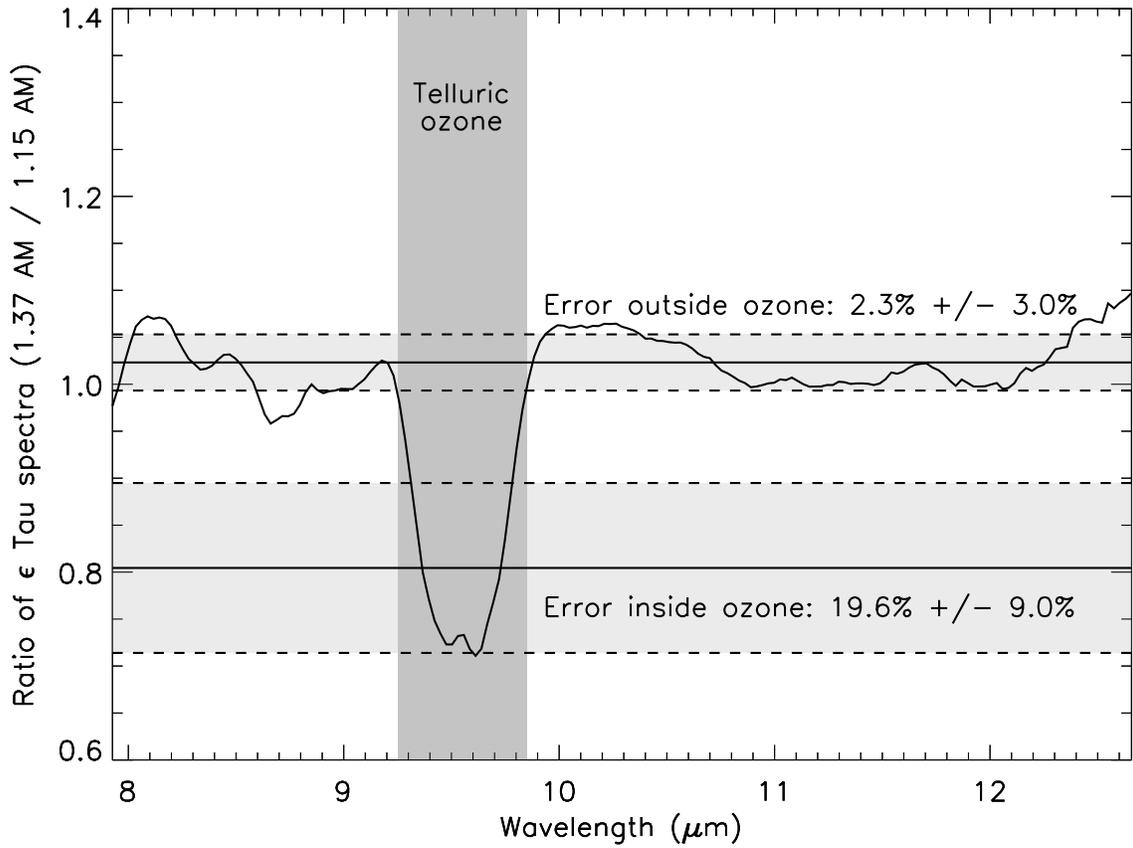}
\caption{The divided, processed spectra of two observations of $\epsilon$ Tau demonstrates the quality of our absolute calibration.  After a telluric airmass correction (described in text), we are able to achieve a calibration discrepancy of 2.3\% $\pm$ 3.0\% outside of the grey-boxed region coincident with telluric ozone.  Inside ozone the discrepancy is 19.6\% $\pm$9.0\%.
\label{Eps_Tau_ratio}}
\end{figure}

All three groups of spectroscopic observations were reduced with our custom artifact removal software \citep{2008ApJ...676.1082S} and mean combined.  We compared the two reduced $\epsilon$ Tau spectra to determine the quality of our flux calibration.  Because the airmass of $\epsilon$ Tau changed between our observations (see Table \ref{All Observations}), we do a telluric airmass correction by dividing our spectra by atmospheric transmission curves for Mauna Kea at the appropriate airmasses \citep[we assumed a 3 mm water vapor column density but using different values did not significantly change our results; the transmission curves are courtesy of Gemini\footnote{http://www.gemini.edu/sciops/telescopes-and-sites/observing-condition-constraints/transmission-spectra} and use the ATRAN code described in][]{ATRAN}.  This technique has also been described in \citet{2004ApJ...601..577H} and \citet{McCabe} among others.  After telluric airmass correction, the ratio of the divided $\epsilon$ Tau spectra is shown in Figure \ref{Eps_Tau_ratio}, boxcar-smoothed with a 0.19$\micron$ kernal.  The large feature centered at $\sim$9.5$\micron$ is telluric ozone, and it is clear that within the grey-boxed region, our flux calibration is poor.  We measure the mean and standard deviation of the residuals and find that outside the ozone feature, our flux calibration has an error of 2.3\% $\pm$ 3.0\%.  The mean difference between the spectra (2.3\%) is treated as a global error, and the standard deviation (3.0\%), while not completely uncorrelated with wavelength, is treated as a local random error.  The same calculation yields an error of 19.6\% $\pm$ 9.0\% error inside ozone.  This could be caused by an incorrect assumption of ozone density in our telluric correction (ozone varies with latitude and season) or a density fluctuation of ozone during our observations (fast day/night or temperature/pressure induced fluctuations).  The best way to address both of these problems is to group future observations closely in both airmass and time.

\begin{figure}
 \includegraphics[angle=0,width=\columnwidth]{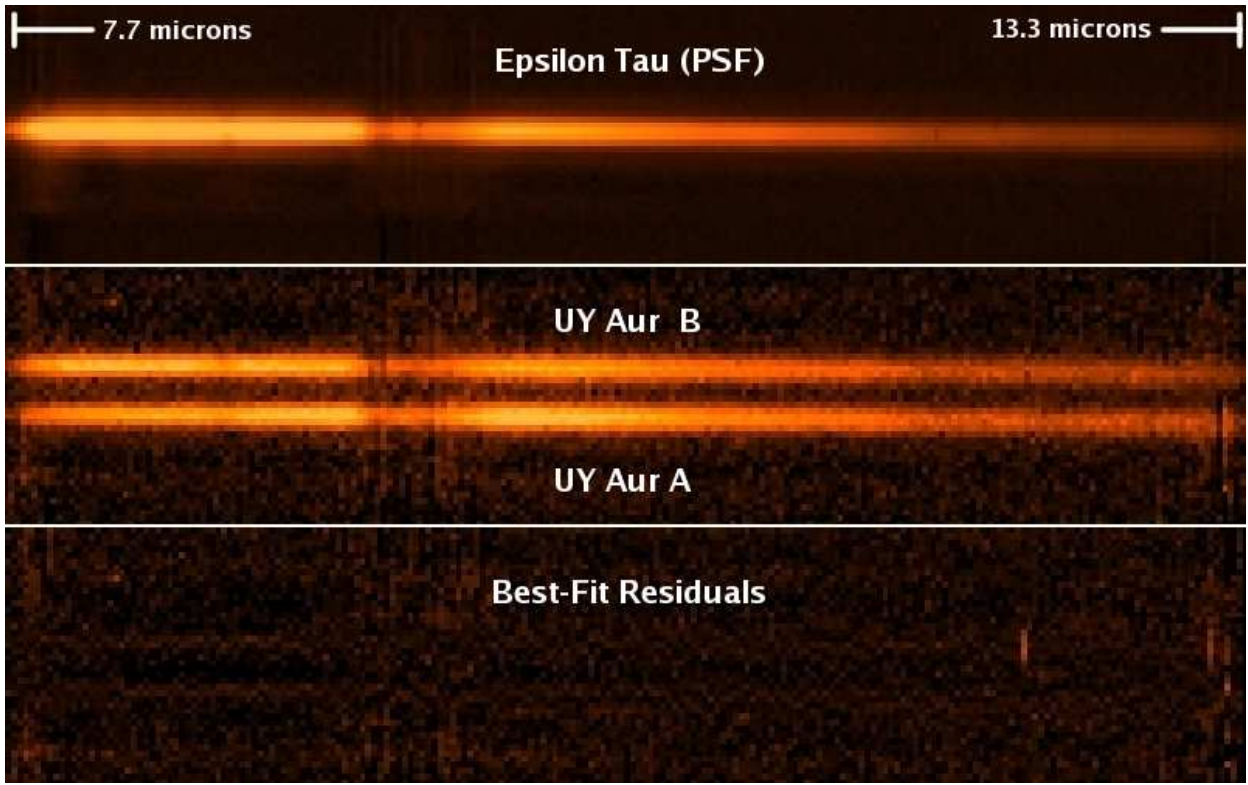}
\caption{Spectra of our PSF/flux calibrator, $\epsilon$ Tau, our science target, UY Aur, and the best-fit residuals after fitting for the flux of each component of UY Aur.  Note: For clarity, the $\epsilon$ Tau spectrum is presented at 1/10 the stretch of the UY Aur spectra and residuals.  UY Aur A is 0.88" from UY Aur B.
\label{MIRAC spectra}}
\end{figure}

Our UY Aur observations were reduced similarly to those of $\epsilon$ Tau.  At a separation of 0.88" \citep{1998ApJ...499..883C, McCabe}, the binary is easily resolved by MMTAO/BLINC-MIRAC4, which has superresolved binaries as tight as 0.12" \citep{2008ApJ...676.1082S}.  We determine the relative flux of the two components of UY Aur at every wavelength by fitting each 1-D slice with a corresponding 1-D slice from the second (closest in airmass and time) observation of the point source $\epsilon$ Tau.  The images of the spectra of $\epsilon$ Tau and UY Aur as well as a best-fit residual plot are shown in Figure \ref{MIRAC spectra}.

We perform an absolute calibration on UY Aur A and B by comparing them to the G9.5III mid-IR standard, $\epsilon$ Tau \citep{Cohen} and using the telluric airmass correction technique described above in this section.  The error analysis on our calibrated spectra is somewhat complex due to the different sources and correlations between errors.  We list the five major error sources here:

\begin{description}

\item[measurement error] due to photon noise, slight PSF mismatch, and crosstalk artifacts on the MIRAC4 chip.  The crosstalk artifacts manifest in our flux measurements as occasional bad measurements at a single wavelength (detector row).  To robustly measure the flux of our spectra in a way that removes the effects of an occasional bad single-row measurement, we median combine our data in 7-pixel (0.19$\micron$) bins (which effectively degrades our spectral resolution by a factor of two).  We empirically calculate the dispersion in our median measurements by taking the separation between the second and sixth ordered values in each 7-pixel bin, and multiplying by the Monte Carlo simulation determined value, 0.303, which gives a 1-sigma (gaussian) error bar on the median.  Typical errors range from 1\%-4\% of the flux with errors up to $\sim$5\% of the flux inside telluric ozone.

\item[global model error] (i.e. correlated throughout the full spectrum) on the \citet{Cohen} model of $\epsilon$ Tau is 2.259\% throughout the spectrum.

\item[local model error] on the \citet{Cohen} model of $\epsilon$ Tau ranges from 1.032\%-1.265\% across the spectral range.

\item[global telluric error] based on our comparison of the two $\epsilon$ Tau measurements shown in Figure \ref{Eps_Tau_ratio}.  We assume a global error of 2.3\% outside the ozone feature and 10\% inside ozone.  The 10\% error inside ozone is less than the 19.6\% discrepancy shown in Figure \ref{Eps_Tau_ratio}.  However, we believe 10\% to be more appropriate based on the fact that the second $\epsilon$ Tau measurement and UY Aur measurement from Table \ref{All Observations} are separated by less in airmass/time than the two $\epsilon$ Tau measurements.

\item[local telluric error] based on our comparison of the two $\epsilon$ Tau measurements shown in Figure \ref{Eps_Tau_ratio}.  We assume errors of 3.0\% outside the ozone feature and 5\% inside ozone.  Again, the 5\% error inside ozone is less than the 9.0\% discrepancy shown in Figure \ref{Eps_Tau_ratio}, but is reasonable considering the proximity of our observations.

\end{description}

\begin{figure}
 \includegraphics[angle=90,width=\columnwidth]{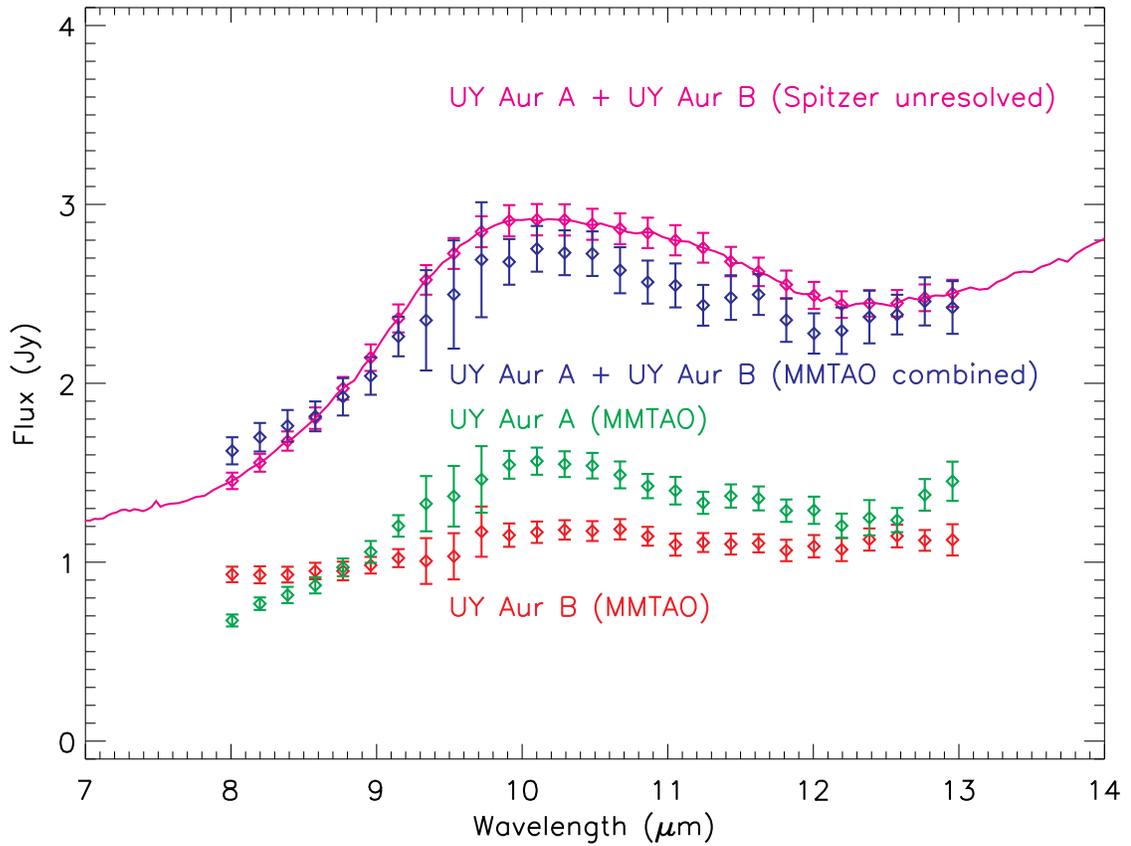}
\caption{MMTAO/BLINC-MIRAC4 flux calibrated spectra of UY Aur A, UY Aur B, their combined spectrum and a comparison Spitzer (unresolved) spectrum.  The error bars shown are the combination of several error sources (described in text).  
\label{UY Aur spectra}}
\end{figure}

All five error sources are combined in quadrature and plotted about the median binned spectra in Figure \ref{UY Aur spectra}.  The figure shows four spectra: UY Aur A, UY Aur B, the combined UY Aur A and B, and a reference Spitzer/IRS spectrum of the unresolved UY Aur binary (Jeroen Bouwman and Ilaria Pascucci, private communication).  Measurement errors (and total errors) are calculated separately for each spectrum.  The Spitzer/IRS spectrum uses Spitzer pipeline S18.7.0 and methods described in \citet{Bouwman2008}, as well as a correction for pointing offest errors using the methods described in \citet{Swain}.  The Spitzer/IRS spectrum has been binned to the same wavelength range as the MMTAO/BLINC-MIRAC4 data, and measurement errors are calculated similarly to the MMTAO/BLINC-MIRAC4 data to include the ``error" resulting from binning a non-neglible spectral range.  The dominant source of error for Spitzer is a 3\% global model uncertainty.

The Spitzer/IRS spectrum has slightly more flux than the MMTAO/BLINC-MIRAC4 UY Aur combined spectrum.  However, the difference is not significant enough to infer variability based on the overall flux levels.  In Section 4, we will show that the shape of the spectra and the relative fluxes of the binary components suggest temporal variability.  It is also evident from this plot that UY Aur A and UY Aur B have similar overall fluxes but very different silicate profiles.  This will be discussed in Section 4.1 when we analyze the shapes of the silicate features.  The flux ratio of the two components is plotted as a function of wavelength in Figure \ref{UY Aur ratio}.  Because relative photometry has fewer error sources than absolute photometry, flux ratios are powerful measurements for determining differences between the silicate feature (Section 4.1) and variability (Section 4.2).

\subsection{Gemini/Michelle Imaging}
\begin{figure}
 \includegraphics[angle=0,width=\columnwidth]{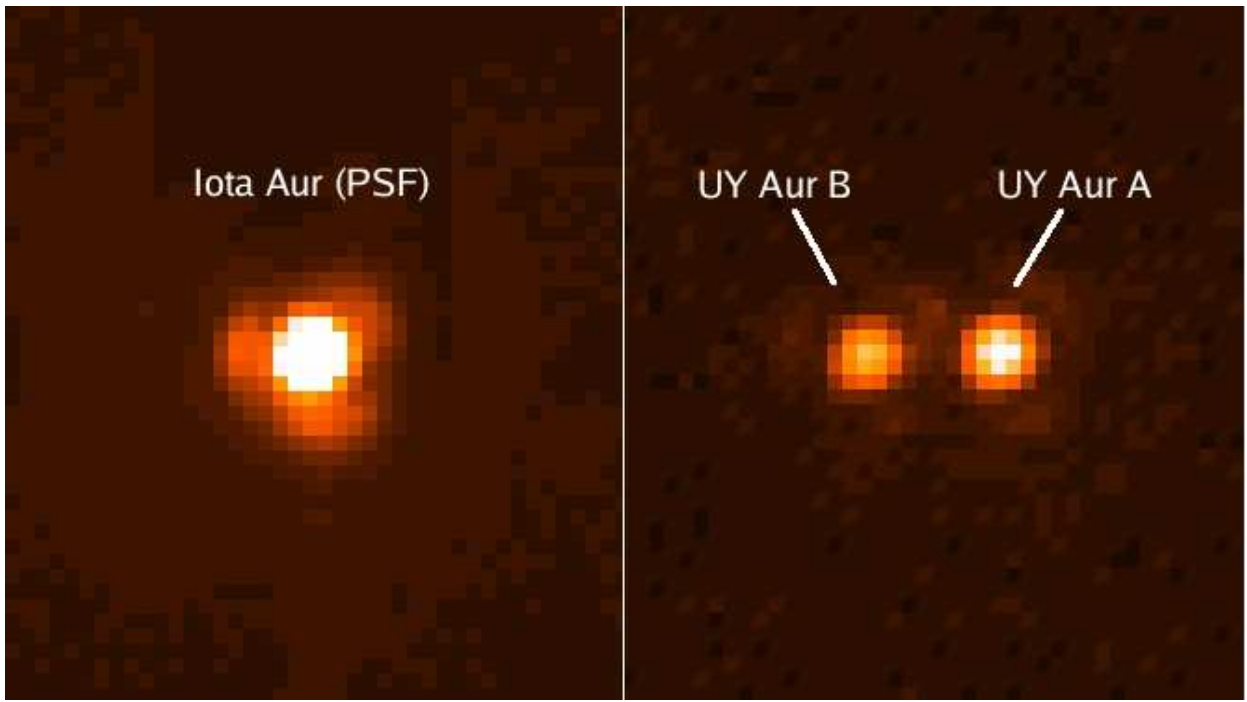}
\caption{Gemini/Michelle acquisition images of Iota Aur (the PSF/flux calibrator) and UY Aur.  For Iota Aur, North is up and East is left.  For UY Aur, the coordinate system has been rotated clockwise by 132 degrees.
\label{Gemini images}}
\end{figure}

On September 27, 2004 UT, Gemini/Michelle observed UY Aur using Michelle's spectroscopic mode for Gemini program GN-2004B-Q59.  We obtained acquisition images of UY Aur (which is easily resolved) and the standard star $\iota$ Aur in the Si-5 filter (11.15$\micron$-12.25$\micron$ at half-maximum).  A summary of these observations is presented in Table \ref{All Observations}.  We did not attempt to use the spectroscopic data as the binary was aligned perpendicular to Michelle's slit. The data were reduced with the standard Gemini IRAF pipeline and are shown in Figure \ref{Gemini images}.

We performed photometry on UY Aur with IRAF \textit{daophot} \citep{daophot} using $\iota$ Aur as a PSF and photometric standard.  $\iota$ Aur is a K3II Cohen mid-IR standard \citep{Cohen, cohen2001}, which has been flux-calibrated to 63.074 Jy in the Si-5 filter\footnote{http://www.gemini.edu/sciops/instruments/mir/MIRStdFluxes.txt}.  The error on our absolute flux calibration is dominated by variable sky conditions, since $\iota$ Aur and UY Aur were observed 15 minutes apart.  Based on the well-matched airmass and weather conditions, and the fact that the data were taken on a relatively dry night in a clear region of the N-band window, we assign a 5\% absolute photometric error to our measurements.  The uncertainty in our UY Aur A-B flux ratio measurement is most likely dominated by PSF fitting error.  Between the two measurments of UY Aur, the flux ratio varied by $\sim$1\%.  We conservatively adopt 2\% as the error on flux ratio.  The results of our Gemini/Michelle imaging photometry are presented in Table \ref{Photometry}.

\subsection{IRTF/MIRAC2 Imaging} 
\begin{figure}
 \includegraphics[angle=0,width=\columnwidth]{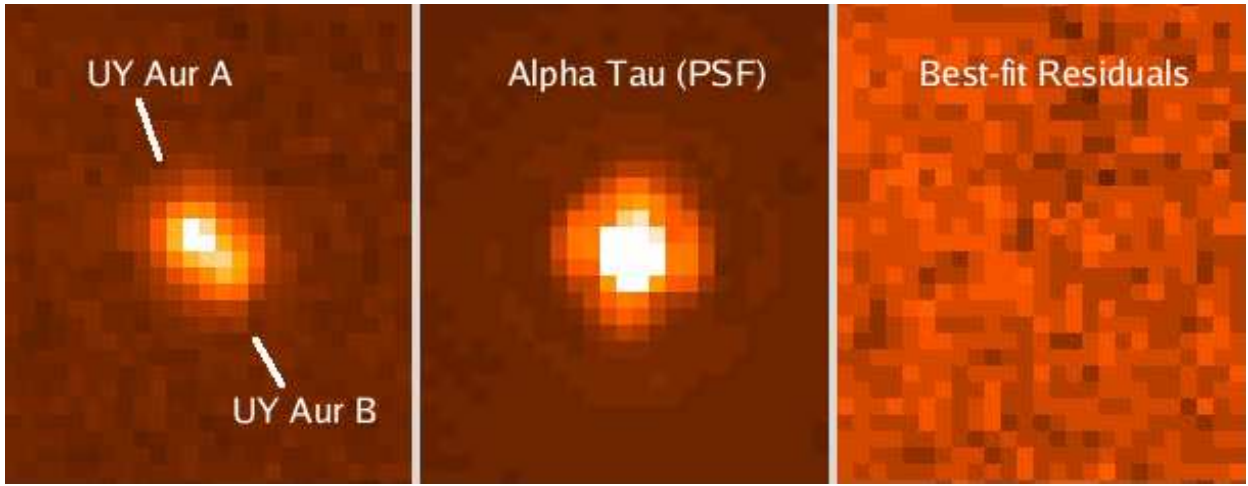}
\caption{IRTF/MIRAC2 images of UY Aur and Alpha Tau (the PSF/flux-calibrator) along with the best-fit residuals of the PSF-fitting photometry.  The best-fit residuals are at the photon noise floor of our image, which gives confidence that the binary was accurately resolved and extracted.  For all three images, North is up and East is left.
\label{IRTF images}}
\end{figure}

On November 15, 1998 UT, IRTF/MIRAC2 observed UY Aur using MIRAC2's imaging mode with the older (now replaced) 128x128 detector.  We obtained images of UY Aur (which is marginally resolved) and the standard star $\alpha$ Tau in the 10.3$\micron$ filter (9.73$\micron$-10.73$\micron$ at half-maximum).  A summary of these observations is presented in Table \ref{All Observations}. The data were reduced with the custom IRAF pipeline of \citet{2003ApJ...587..407C} and are shown in Figure \ref{IRTF images}.

We performed PSF fitting photometry on UY Aur with IRAF \textit{daophot} \citep{daophot} using $\alpha$ Tau as a PSF and photometric standard.  $\alpha$ Tau is a MIRAC2 standard, which has been flux-calibrated to 584 Jy in the 10.3 $\mu$m filter\footnote{cfa-www.harvard.edu/\textasciitilde jhora/mirac/mrcman.pdf} using the \citet{cohen92} system defined by $\alpha$ Lyr. The night was photometric and both the object and the standard were observed at very low airmass (1.02-1.09).  However, the 10.3$\micron$ filter partially overlaps with the telluric ozone feature (which can adversely effect absolute calibrations) so we conservatively assume an absolute calibration error of 8\%, which is added in quadrature to the photon/fitting errors.
We also adopt 4\% as the error on the flux ratio from the errors given by the IRAF {\it allstar} task. The results of our IRTF/MIRAC2 imaging photometry are presented in Table \ref{Photometry}.

\section{Analysis}
\subsection{Normalization of the Spectra}
\begin{figure}
 \includegraphics[angle=90,width=6.0in]{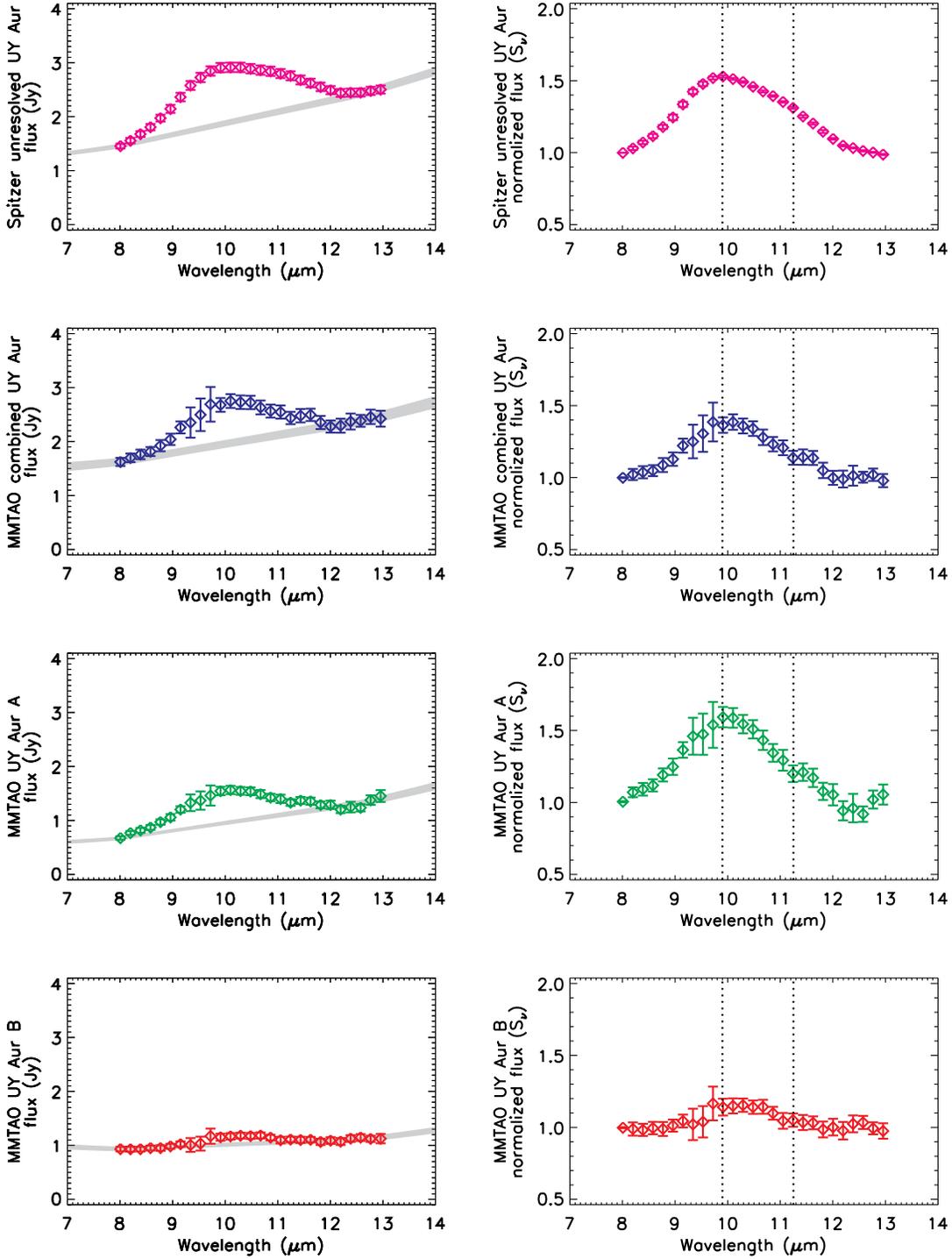}
\caption{LEFT: MMTAO/BLINC-MIRAC4 spectra shown in Figure \ref{UY Aur spectra} with the 1-$\sigma$ range of our Monte Carlo continuum plotted in grey.  RIGHT: Normalized spectra (see Equation \ref{eq:normalization_form}) after continuum subtraction.  The error bars are the result of our Monte Carlo analysis (described in Section 3.1).
\label{UY Aur normalized}}
\end{figure}

Following the example of numerous authors \citep{2003A&A...400L..21V,2003A&A...412L..43P,2006ApJ...639..275K,Ilaria}, we continuum subtract and normalize our spectra (see Figure \ref{UY Aur normalized}) in order to measure size/shape parameters of the 10$\micron$ silicate feature, which can be used to infer the characteristic dust grain size, and crystallinity.  After continuum subtraction, the normalization follows the formula 
\begin{equation}\label{eq:normalization_form}
{S_{\nu}=1+\frac{F_{\nu}-F_{\nu,c}}{<F_{\nu,c}>}}
\end{equation}
where $F_{\nu}$ is the spectrum and $F_{\nu,c}$ is the fit continuum and the denominator, $<F_{\nu,c}>$ is the continuum averaged over the wavelength range of the silicate feature.

\begin{figure}
 \includegraphics[angle=90,width=6.0in]{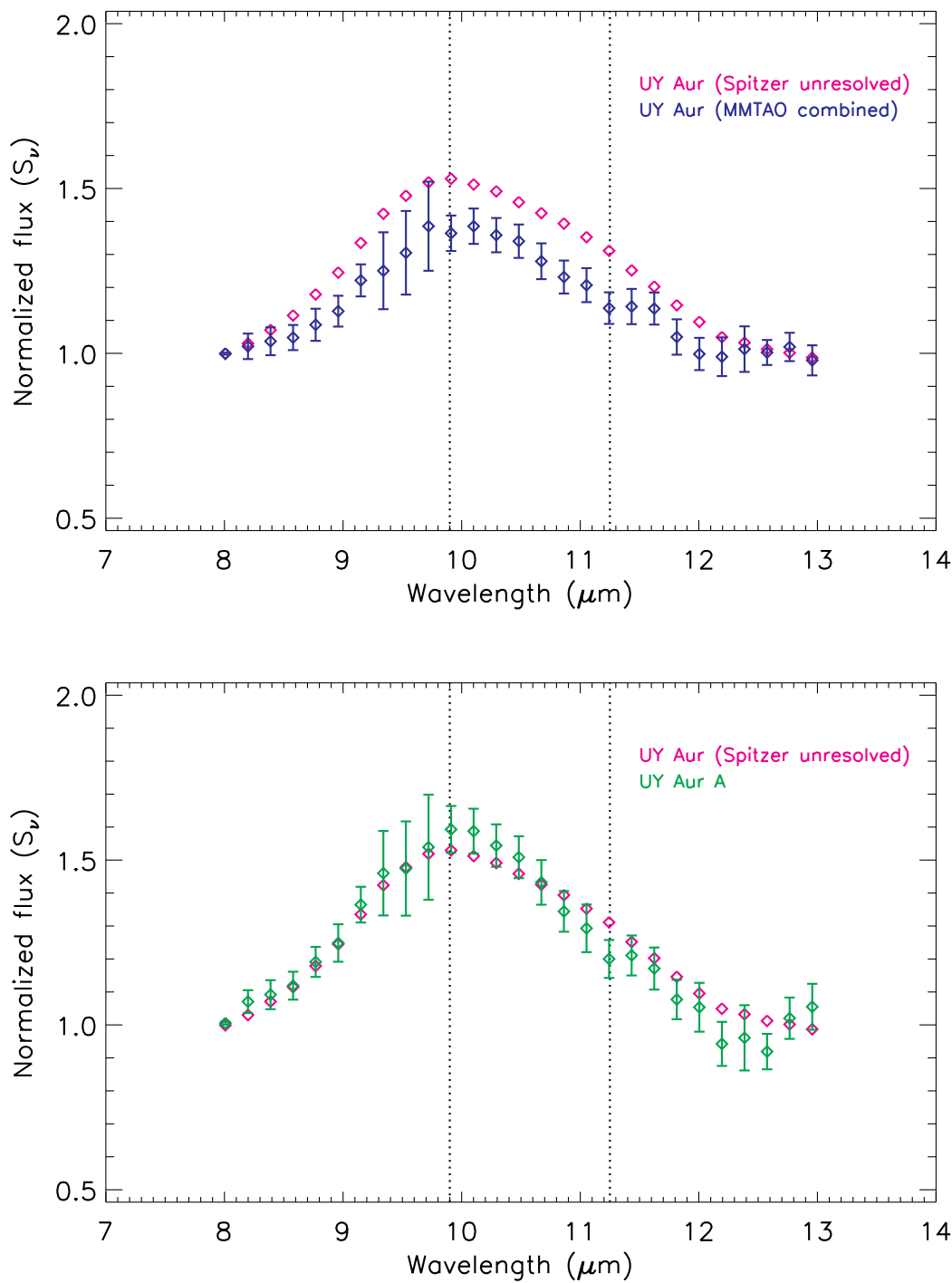}
\caption{Comparisons of normalized spectra from Figure \ref{UY Aur normalized}.  The negligible error bars on the Spitzer normalized spectra (shown in Figure \ref{UY Aur normalized}) are removed for clarity.  Our combined UY Aur normalized spectrum shows statistically significant variability compared to the Spitzer spectrum.  However, UY Aur A itself is statistically indistinguishable from the Spitzer spectrum. 
\label{UY Aur normalized comparison}}
\end{figure}

We perform a Monte Carlo simulation of the error terms described in Section 2.1 to construct 10000 sets of synthetic spectra.  The error terms are correlated where appropriate so that errors from the absolute calibration (global and local model errors and global and local telluric errors) are the same for the three MMTAO/BLINC-MIRAC4 spectra (UY Aur A, UY Aur B, and combined UY Aur) and global model/telluric/calibration uncertainties are the same throughout each full spectrum in a given Monte Carlo trial.  For each trial we perform a best-fit linear continuum subtraction based on the continuum at 8.0$\micron$ and 12.5$\micron$-13.0$\micron$.   Since the N-band atmospheric window precludes us from measuring the quadratic and higher order continuum terms, we adopt the quadratic term of the continuum from the Spitzer/IRS spectrum (fit between 5.3$\micron$-8.0$\micron$ and 12.6$\micron$-14.2$\micron$; to avoid silicate) for all of our spectra.  As can be seen from an independent analysis by \citet{Watson}, the higher order (non-linear) continuum terms for UY Aur are small. The Monte Carlo analysis gives us a range of values for the normalized flux at each wavelength from which we calculate error bars.  \citet{Ilaria} use a similar Monte Carlo analysis (with different error sources).  The 1-$\sigma$ range of the fit continua is shown in Figure \ref{UY Aur normalized} along with the normalized fluxes derived from equation \ref{eq:normalization_form} and our Monte Carlo analysis.  A few of the normalized spectra are also plotted against each other for ease of comparison in Figure \ref{UY Aur normalized comparison}.  Table \ref{shape/strength} shows the 9.9$\micron$ normalized flux and the ratio of the 11.25$\micron$/9.9$\micron$ normalized fluxes, which are used as a proxy for grain-size and crystallinity respectively.  We also measure the ratio of these parameters for UY Aur A and B (since this helps supress much of the uncertainty caused by the absolute error terms).  We find that the ratio of the 9.9$\micron$ normalized flux terms is 1.40$\pm$0.06 and the ratio of the 11.25$\micron$/9.9$\micron$ normalized color is 0.82$\pm$0.04.
Because the quadratic term of our continuum subtraction is not well known (we adopt it from a different epoch and it is not resolved for the two binary components), we may have some residual systematic errors that we do not present. However, the difference between using the quadratic term from the Spitzer data and using no quadratic term at all creates a discrepancy of only 0.01 for the values in Table \ref{shape/strength} so we believe this effect to be small.

\subsection{Photometric Variability}
We compare the photometry of multiple epochs of UY Aur observations to check for photometric variability, presenting both absolute photometry of the full system and relative photometry of the resolved binary for the ground-based data in Table \ref{Photometry}.  With the spectroscopic data, we can construct synthetic filters to compare our actual photometry (IRTF/MIRAC2, Keck/OSCIR, Gemini/Michelle) to the spectra (MMTAO/BLINC-MIRAC4 and Spitzer/IRS).  For example, photometry in the Gemini/Michelle Si-5 filter is compared to MMTAO/BLINC-MIRAC4 spectra that have been convolved onto the Gemini/Michelle Si-5 photometric system.  This necessitates a thorough understanding and calibration of each photometric system.

The traditionally quoted energy flux of an object through a photometric system is given by
\begin{equation}\label{eq:absolute lambda}
F(filter)=\frac{\int F_{\lambda}(\lambda) \cdot R(\lambda) d\lambda}{\int R(\lambda) d\lambda}
\end{equation}
where $F_{\lambda}(\lambda)$ is the spectrum of the object as a function of $\lambda$ and $R(\lambda)$ is the system response function (i.e. the product of the sky transmission, the system QE and the filter profile as a function of $\lambda$).  If we convert this to an $F_{\nu}$ system (i.e. Janskys), the equation becomes
\begin{equation}\label{eq:absolute nu}
F(filter)=\frac{\int F_{\nu}(\lambda) \cdot c \cdot \lambda^{-2} \cdot R(\lambda) d\lambda}{\int R(\lambda) d\lambda}
\end{equation}
However, when comparing detector counts for relative and absolute calibration, it is necessary to integrate photons rather than energy \citep{MSBessell}, which necessitates that we divide the flux term in equation \ref{eq:absolute nu} by $E=h\nu$.  This gives us
\begin{equation}\label{eq:relative photometry}
\frac{Counts_{1}}{Counts_{2}}=\frac{\int F_{\nu, 1}(\lambda) \cdot \lambda^{-1} \cdot R(\lambda) d\lambda}{\int F_{\nu, 2}(\lambda) \cdot \lambda^{-1} \cdot R(\lambda) d\lambda}
\end{equation}
where $\frac{counts_{1}}{counts_{2}}$ is the ratio of photons detected for object 1 compared to object 2, with fluxes $F_{\nu,1}$ and $F_{\nu,2}$ respectively.

In this formalism, Equations \ref{eq:absolute lambda} or \ref{eq:absolute nu} should be used to determine absolute fluxes for photometry (i.e. to put a Cohen model spectrum onto a photometric system) while Equation \ref{eq:relative photometry} should be used for photometry with actual data (i.e. counts measured for object 1 versus counts measured for object 2 in a photometric system).

In order to put the Spitzer/IRS and MMTAO/BLINC-MIRAC4 spectra onto the photometric systems of IRTF/MIRAC2, Keck/OSCIR and Gemini/Michelle, we use Equation \ref{eq:relative photometry} with \citet{cohen92,1992AJ....104.2030C,cohen95,Cohen} spectra of $\alpha$ Tau (for IRTF/MIRAC2), $\alpha$ Lyr (for Keck/OSCIR) and $\iota$ Aur (for Gemini/Michelle) as $F_{\nu, 2}(\lambda)$.  We use the telluric absorption spectrum at 1.0 airmasses for Mauna Kea assuming 1.0 mm precipitable water vapor as described in Section 2.1 \citep{ATRAN}, and we use filter curves for 10.3$\micron$\footnote{http://www.cfa.harvard.edu/\textasciitilde emamajek/mirac/index.html} (IRTF/MIRAC2), N-band\footnote{http://www.gemini.edu/sciops/instruments/oscir/oscirIndex.html} (Keck/OSCIR) and Si-5\footnote{http://www.gemini.edu/sciops/instruments/michelle/imaging/filters} (Gemini/Michelle).

Equation \ref{eq:relative photometry} and its corresponding uncertainty is calculated using a Monte Carlo approach with the same error sources (on $F_{\nu,1}$ and $F_{\nu,2}$) and functionality as the analysis described in Section 3.1.  The binary flux ratios and corresponding uncertainties are calculated using the relative errors shown in Figure \ref{UY Aur ratio}.

\section{Discussion}
\subsection{Normalized Spectra}
Our continuum-subtracted, normalized spectra in Figure \ref{UY Aur normalized} show that UY Aur A's silicate emission mostly dominates the combined normalized spectrum, but that UY Aur B's role is not insignificant.  A direct comparison of the MMTAO/BLINC-MIRAC4 normalized spectrum (Figure \ref{UY Aur normalized comparison}) with the Spitzer normalized spectrum shows that the feature itself has changed, as the Spitzer spectrum has a stronger feature that could be interpreted as smaller, less-evolved grains.  Interestingly, the bottom panel shows that the MMTAO/BLINC-MIRAC4 normalized spectrum of UY Aur A is consistent with the Spitzer normalized spectrum of the unresolved pair.  This could mean that UY Aur A's feature was even more dominant at the time of the Spitzer observations.  Given that we have observed significant variability in both binary components (see below in Section 4.2), it is possible that changes in either star's silicate feature could have caused this.  Without two epochs of resolved spectra (the Spitzer spectrum is a blend), it is impossible to know if UY Aur A's silicate feature has varied, or just its continuum.

\begin{figure}
 \includegraphics[angle=90,width=\columnwidth]{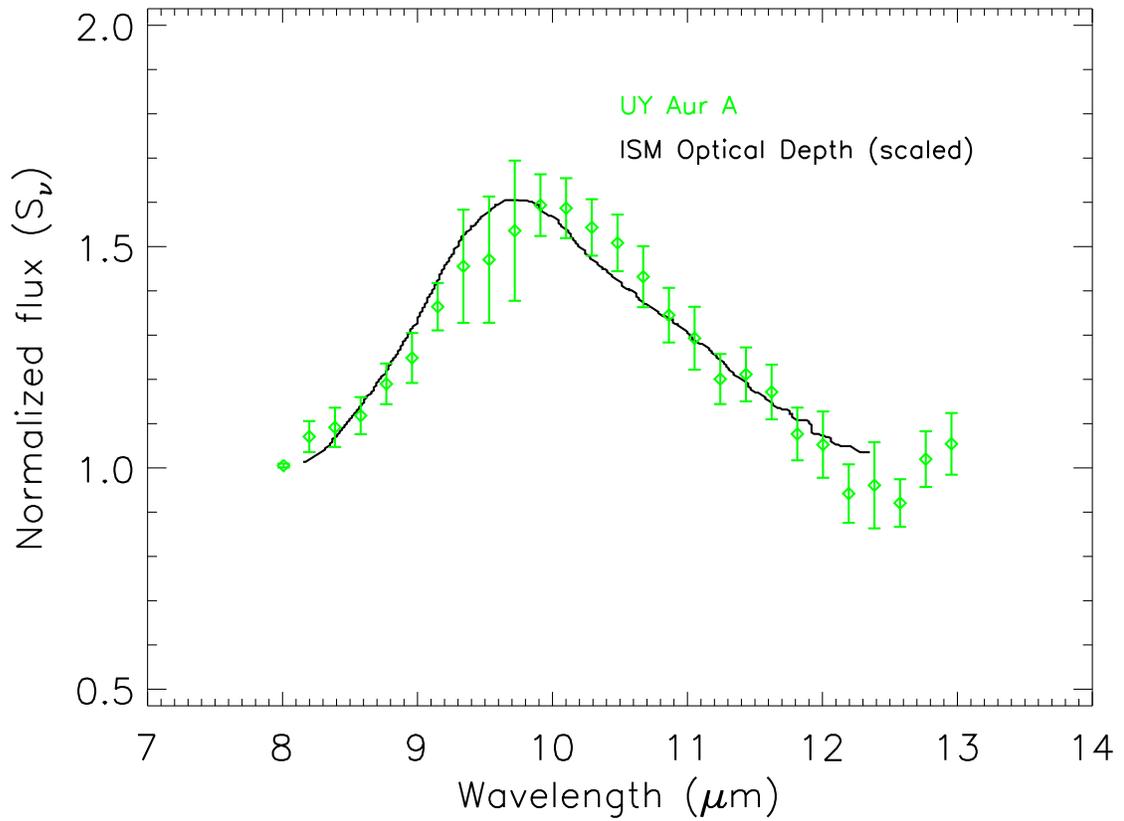}
\caption{The continuum-subtracted, normalized spectrum of UY Aur A is compared to the (scaled) optical depth of the interstellar medium towards Sgr A*.  The curves are consistent with reduced $\chi^{2}=0.75$.  The fact that UY Aur A's dust is very similar to the ISM either means that it has not evolved (via grain-growth), or that variable processes (such as accretion from the circumbinary disk) are replenishing the small $\sim$0.1$\micron$ grains.
\label{galactic center}}
\end{figure}

As was noted by \citet{Watson}, the Spitzer spectrum of the unresolved UY Aur binary shows almost uniquely clear evidence for small, pristine, ISM-like grains.  In Figure \ref{galactic center}, we compare our normalized spectrum of UY Aur A with the scaled optical depth of the ISM towards Sgr A* using the measurements of \citet{2004ApJ...609..826K} who find that the ISM is dominated by $\sim$0.1$\micron$ amorphous grains.  Our spectrum is consistent with these ISM grains with a reduced $\chi_{\nu}^{2}$ value of 0.75.  Given that UY Aur A is variable (see below in section 4.2) and has pristine ISM-like grains, it will be interesting to see if future resolved spectroscopy finds a varying silicate feature similar to the recent observations of EX Lupi \citep{2009Natur.459..224A}.

\begin{figure}
 \includegraphics[angle=90,width=\columnwidth]{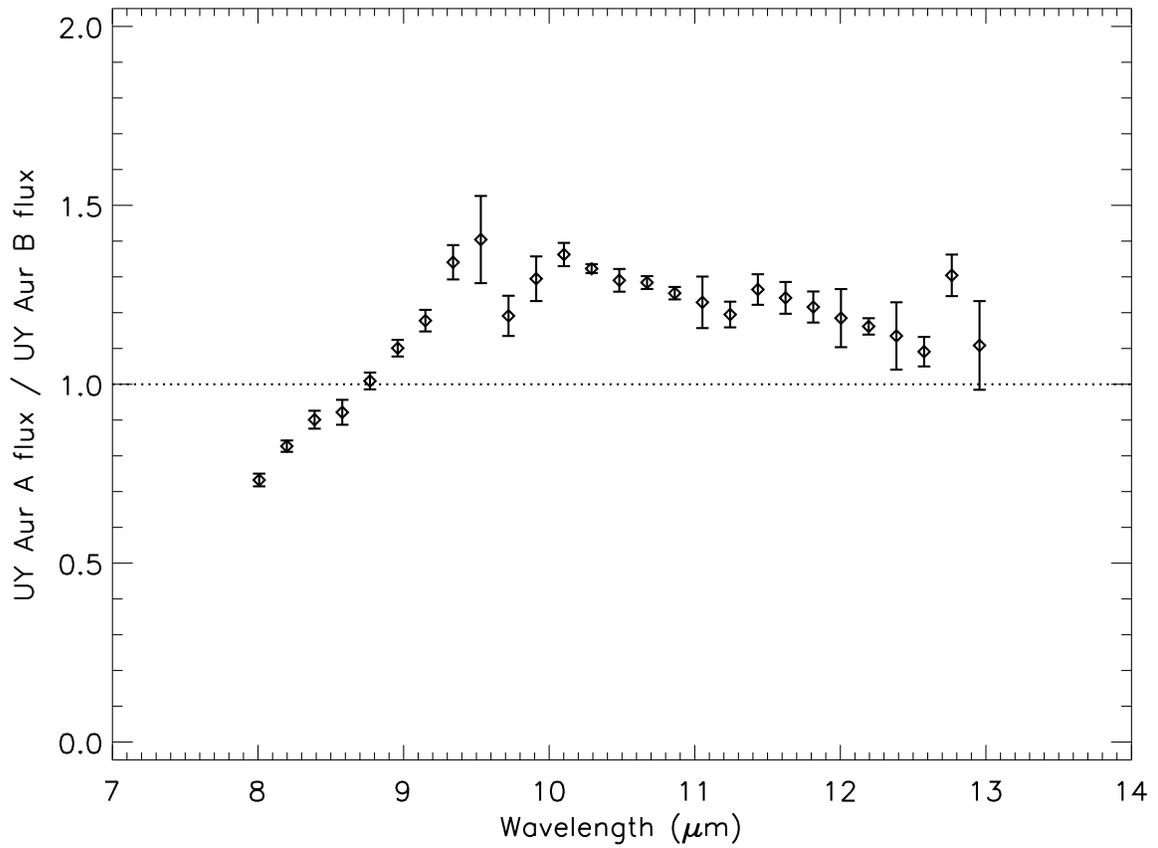}
\caption{The flux ratio of UY Aur as a function of wavelength from the MMTAO/BLINC-MIRAC4 spectra.  Because the relative errors on each components flux are significantly lower than the absolute flux calibration errors,  flux ratios provide a powerful measurement for demonstrating variability and differences between the silicate features.
\label{UY Aur ratio}}
\end{figure}

\begin{figure}
 \includegraphics[angle=90,width=\columnwidth]{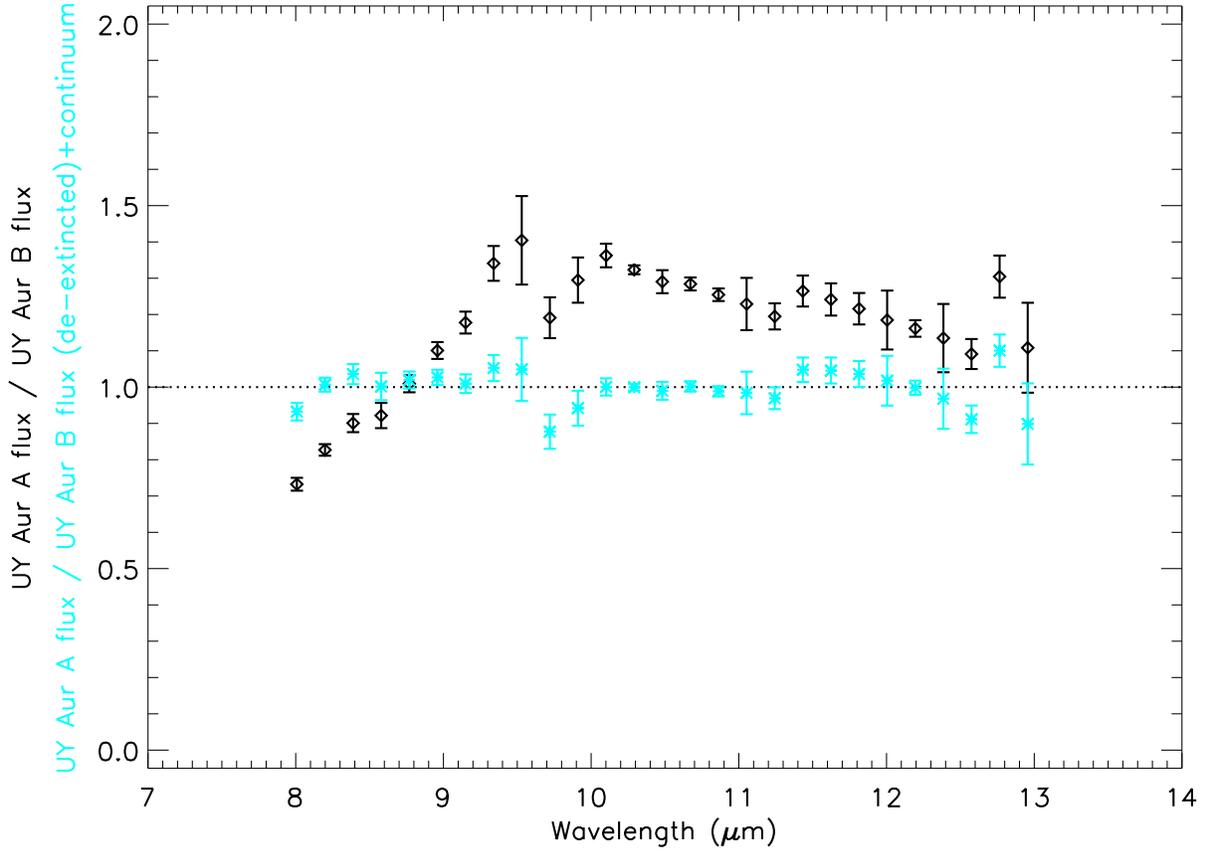}
\caption{In order to see if UY Aur B's silicate feature is similar to UY Aur A's but with more extinction and a different underlying continuum, we take the observed relative fluxes between UY Aur A and B and do a best-fit de-extinction of B, plus the addition of a linear continuum, to try to produce an equal flux-ratio across all wavelengths.  The observed relative fluxes between UY Aur A and B are shown as black diamonds (as in Figure \ref{UY Aur ratio}).  The best-fit de-extincted ratio (+continuum) is shown as teal asterisks and has a reduced $\chi_{\nu}^{2}$ of 1.71 with respect to the dashed line (unity).
\label{ratio de-extincted}}
\end{figure}

\begin{figure}
 \includegraphics[angle=90,width=3.5in]{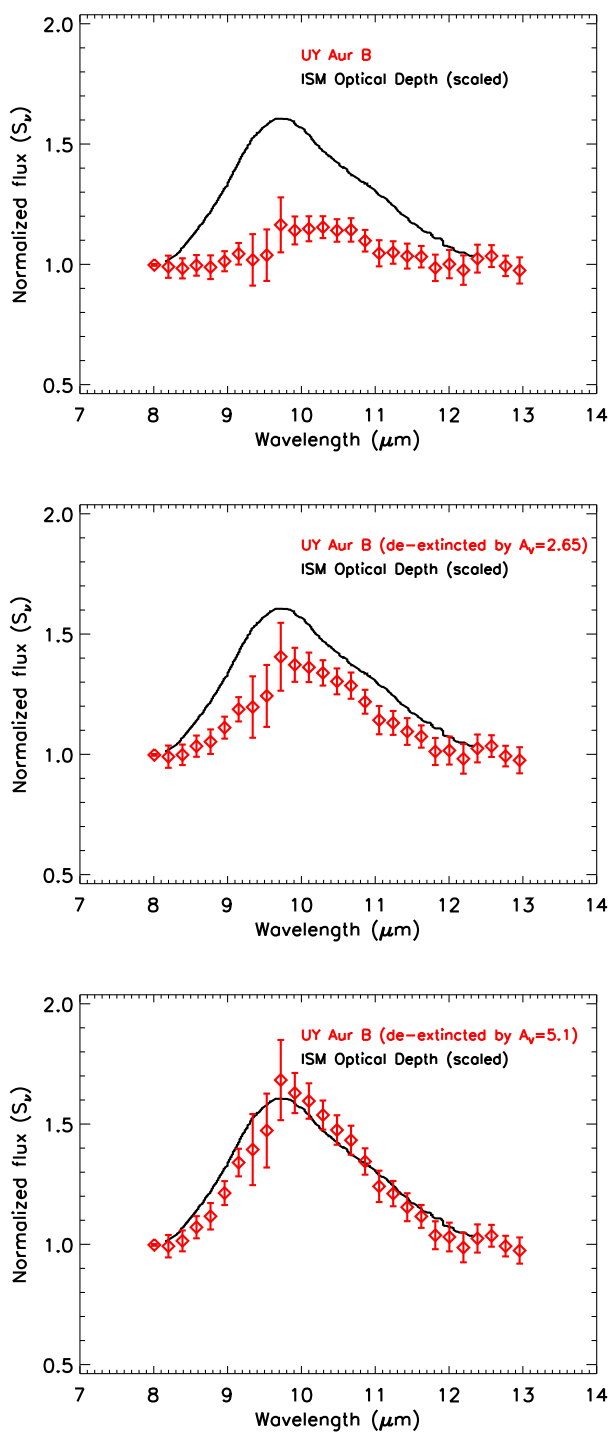}
\caption{Normalized spectra of UY Aur B after de-extinction compared to the scaled ISM-optical depth shown in Figure \ref{galactic center}.  From top to bottom, UY Aur B has been de-extincted by $A_{V}$ of 0, 2.65 \citep[the value measured by][]{2003ApJ...583..334H} and 5.1 (the best-fit value to make UY Aur B's silicate feature like UY Aur A's).  Assuming a large amount of extinction towards UY Aur B, it is possible that UY Aur B's silicate grains are similar to UY Aur A's.  For lower values of extinction, UY Aur B's silicate grains would have to be larger than UY Aur A's grains.
\label{UY Aur B de-extincted}}
\end{figure}

While it would be interesting to compare the dust properties of UY Aur A and B, extinction in the spectrum of B precludes a direct comparison.  Instead, we can de-extinct the spectrum of UY Aur B, to see if we can reproduce UY Aur A's ISM-like silicate feature with a plausible amount of extinction.  We use the ratio of the fluxes between UY Aur A and B (Figure \ref{UY Aur ratio}, which has lower errors than our flux-calibrated spectra) and de-extinct UY Aur B using scaled ISM optical depths from \citet{2004ApJ...609..826K}.  These scaled optical depths can be converted to $A_{V}$ using the ISM extinction curve of \citet{1985ApJ...288..618R}.  With this, we do a 3 parameter best-fit ($A_{V}$ and a linear continuum) to match the spectrum of UY Aur B to UY Aur A.  Our results are plotted in Figure \ref{ratio de-extincted} where the black diamonds are the observed flux ratios between UY Aur A and B (Figure \ref{UY Aur ratio}), and the teal asterisks are the theoretical flux ratios between UY Aur A and B after a best-fit de-extinction of UY Aur B plus the addition of a linear continuum.  The best-fit extinction value is $A_{V}=5.1$ with a reduced $\chi_{\nu}^{2}$ of 1.71.  This is a plausible amount of extinction given that UY Aur B has been observed to vary by $>$4.5 magnitudes at R-band, due to a mixture of variable extinction and accretion \citep{1995ApJ...444L..93H}.  Figure \ref{UY Aur B de-extincted} shows the normalized spectrum of UY Aur B after being de-extincted by $A_{V}=$ 0 (as observed), 2.65 \citep[the value measured by][]{2003ApJ...583..334H} and 5.1 (the best-fit value to make UY Aur B's silicate feature like UY Aur A's).  If extinction in UY Aur B is variable, then 5.1 magnitudes of $A_{V}$ extinction would allow the dust grains in UY Aur A and B to be similar.  If extinction in UY Aur B is not variable or was lower than $A_{V}$=5.1 at the time of our observations, as was true when measured by \citet{2003ApJ...583..334H}, then the silicate feature in UY Aur B is shallower than A's which would mean it has larger grains.  Disentangling these two scenarios is only possible with simultaneous resolved extinction measurements and resolved mid-IR spectroscopy.

\subsection{Photometric Variability}
\begin{figure}
 \includegraphics[angle=90,width=\columnwidth]{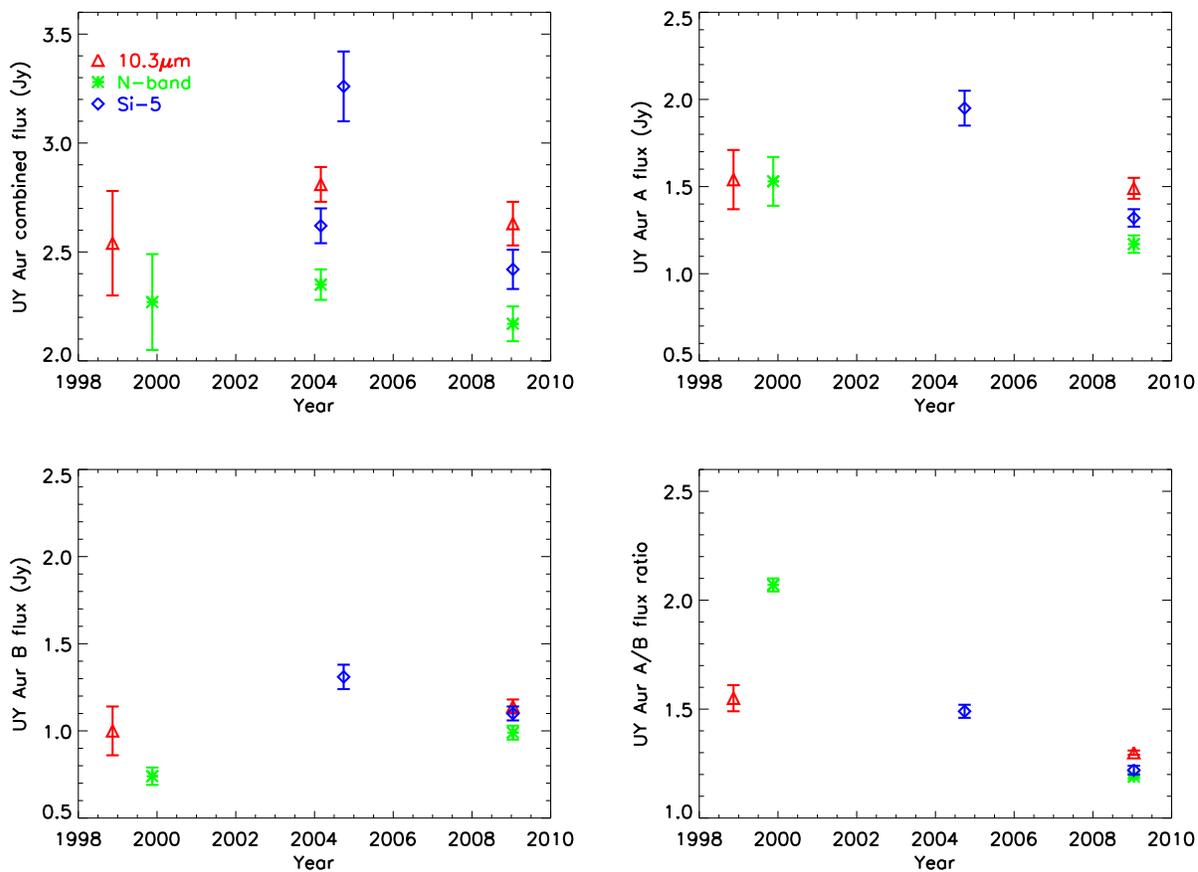}
\caption{Variability of UY Aur and its two components in three filters.  The photometry is listed in Table \ref{Photometry}.  Variability is seen, to some extent, in the combined UY Aur plot, as well as in the UY Aur A and B plots.  However, ground-based resolved data are most sensitive to changes in the flux ratio (seen in the bottom right).  Our most recent MMTAO/BLINC-MIRAC4 data show that the binary is much closer to equal flux than it has been in the past in all three filters.
\label{variability plot}}
\end{figure}

As is evident from Table \ref{Photometry} and a corresponding plot shown in Figure \ref{variability plot}, the UY Aur system is variable in the mid-infrared.  Among our results, we note that UY Aur A and UY Aur B became significantly fainter in the Si-5 filter between September 27, 2004 and January 14, 2009.  Also, between November 16, 1999 and January 14, 2009, UY Aur A got fainter in N-band while UY Aur B got brighter.  This change is further highlighted in the flux ratio measurements which dropped from 2.07$\pm$0.03 to 1.19$\pm$0.01 over that period.  Our results are somewhat surprising considering that UY Aur B was assumed to be the dominant source of variability in the near-infrared \citep{2003AJ....126.2009B}.  The fact that UY Aur A is also variable in the mid-infrared may mean that it deserves a more complete study at shorter wavelengths as well.  Currently, the best resolved near-infrared variability data set, with absolute calibrations, is 3 H-band measurements by \citet{1998ApJ...499..883C} and \citet{2007AJ....134..880H} that show $\sim$0.4 magnitudes of variability for UY Aur A and $\sim$1.3 magnitudes of variability for UY Aur B between 1996 and 2005.  If this variability is caused entirely by extinction, we can use the \citet{1985ApJ...288..618R} ISM extinction law to say that N-band extinction variability would be $\sim$0.1 magnitudes for UY Aur A, and $\sim$0.4 magnitudes for UY Aur B.  Based on our mid-IR variability data from Table \ref{Photometry}, this hypothetical extinction variability would be enough to explain UY Aur B's mid-IR behavior, but not UY Aur A's.  Thus, some other process, such as variable accretion, must be responsible for UY Aur A's mid-IR variability.

If \citet{1998ApJ...499..883C}'s hypothesis \citep[following ][]{1996ApJ...467L..77A} that spiral streamers are responsible for dust-replenishment and variability in the system, then our mid-infrared data suggest the streamers may be affecting the primary star (UY Aur A) as well as the secondary.  Future studies should address whether binaries are more likely than single stars to be variable in the mid-infrared, and whether the variability and the unevolved ISM-like grains are contingent on large circumbinary disks feeding truncated circumstellar disks, such as the ones surrounding UY Aur and GG Tau.  Spectroscopic measurements, which are able to determine if the source of the variability is continuum or optically thin dust emission, would be particularly useful.

\section{Conclusions}
While the Spitzer IRS spectrograph has observed numerous young stellar objects to study grain-growth, Spitzer's 85cm aperture cannot resolve most binaries (tighter than $\sim$3").  As a result, large aperture (ground-based mid-IR systems and/or JWST) must be used to separate tight/medium-separation binaries.

Using the 6.5 meter MMT with MMTAO/BLINC-MIRAC4, we obtained high-Strehl, spatially resolved, low-resolution, N-band spectra of the 0.88" binary UY Aur.  We compare our results to resolved photometry from IRTF/MIRAC2, Gemini/Michelle, and Keck/OSCIR as well as  an unresolved spectrum from Spitzer.  We find the following:

1) UY Aur A and UY Aur B have very different silicate features, which when combined, produce a blended spectrum that includes emission from both sources and absorption from UY Aur B.  Other binaries may have similarly complicated silicate blends and should be resolved by large-aperture telescopes.

2) UY Aur A's silicate spectrum is consistent with dust emission from pristine, ISM-like grains.  UY Aur B, which is an infrared companion, has a much flatter silicate spectrum that may be the result of foreground extinction from an edge-on disk.  As a result, we cannot determine whether UY Aur A and UY Aur B share similar dust properties (although if corrected for $A_{V}$=5.1 magnitudes of extinction, UY Aur B has a similar silicate feature to UY Aur A).  The fact that UY Aur A exhibits ISM-like dust implies that either its dust has not evolved (via grain-growth) or that its small grains are replenished by variable processes (such as streamers from the circumbinary disk reservoir).

3) While UY Aur B has shown optical and near-IR variability in the past, we show that UY Aur A and UY Aur B both show significant variability in the mid-IR.  The variability in UY Aur B is expected based on its presumably changing extinction profile.  However, the mid-IR variability of UY Aur A, especially considering its unevolved, ISM-like grains, is more surprising.  Future resolved spectra will be able to determine if this variability is due to continuum changes, or if it is due to episodes of dust-processing as was recently observed in EX Lupi \citep{2009Natur.459..224A}.

\acknowledgements
The authors are grateful to Jeroen Bouwman and Ilaria Pascucci for supplying the reduced Spitzer spectrum of UY Aur.  We also thank the Gemini mid-IR team, especially Rachel Mason, Scott Fisher, Melanie Clark, and Michael Hoenig for digging up old acquisition images.  We thank Dan Potter and Joe Hora for help with the IRTF observations.  We also thank the anonymous referee for his/her comments, that have greatly improved this paper.  This paper makes use of the IDL library, \textit{mpfit}, described in \citet{mpfit}.  AJS acknowledges the NASA Graduate Student Research Program (GSRP) and the University of Arizona's Technology Research Initiative Fund (TRIF) for their generous support.  LMC's research is supported by NSF CAREER, MRI and TSIP awards.

\clearpage

\begin{deluxetable}{lcccccccccccc}
\tabletypesize{\scriptsize}
\tablecaption{Imaging and Spectroscopic Observations of UY Aur}
\tablewidth{0pt}
\tablehead{
\colhead{Date (UT)} &
\colhead{Telescope/Instrument} &
\colhead{Object} &
\colhead{Filter} &
\colhead{Exposure Length (s)} &
\colhead{\# of Exposures} &
\colhead{Airmass}
}

\startdata

Jan 14, 2009 & MMTAO/BLINC-MIRAC4\tablenotemark{a} & $\epsilon$ Tau & Grism & 10 & 10 & 1.37-1.38 \\
Jan 14, 2009 & MMTAO/BLINC-MIRAC4\tablenotemark{a} & $\epsilon$ Tau & Grism & 10 & 14 & 1.15-1.16 \\
Jan 14, 2009 & MMTAO/BLINC-MIRAC4\tablenotemark{a} & UY Aur         & Grism & 10 & 18 & 1.09-1.11 \\
\hline
Sep 27, 2004 & Gemini/Michelle\tablenotemark{b} & $\iota$ Aur & Si-5 & 4 & 1 & 1.05 \\
Sep 27, 2004 & Gemini/Michelle\tablenotemark{b} & UY Aur      & Si-5 & 4 & 2 & 1.03 \\
\hline
Nov 15, 1998 & IRTF/MIRAC2 & $\alpha$ Tau & 10.3$\micron$ & 5 & 1 & 1.09 \\
Nov 15, 1998 & IRTF/MIRAC2 & UY Aur       & 10.3$\micron$ & 10 & 54 & 1.02-1.05 \\
\enddata
\tablenotetext{a}{All data were taken with MMTAO at a loop-speed of 550 Hz}
\tablenotetext{b}{The data were all taken within a period of 15 minutes and had IQ=70\%, CC=50\% and WV=20\% (Gemini observing constraints for Mauna Kea---www.gemini.edu/sciops/telescopes-and-sites/observing-condition-constraints)}
\label{All Observations}
\end{deluxetable}

\clearpage

\begin{deluxetable}{lcccccccccccc}
\tabletypesize{\scriptsize}
\tablecaption{Photometry of UY Aur}
\tablewidth{0pt}
\tablehead{
\colhead{Date} &
\colhead{Telescope/} &
\colhead{Filter} &
\colhead{UY Aur} &
\colhead{UY Aur A} &
\colhead{UY Aur B} &
\colhead{Binary flux ratio}
\\

\colhead{} &
\colhead{Instrument} &
\colhead{} &
\colhead{Flux (Jy)} &
\colhead{Flux (Jy)} &
\colhead{Flux (Jy)} &
\colhead{(A/B)}

}

\startdata
Nov 15, 1998 & IRTF/MIRAC2 & 10.3$\micron$ & 2.54$\pm$0.24 & 1.54$\pm$0.17 & 1.00$\pm$0.14 & 1.55$\pm$0.06 \\
Feb 27, 2004 & Spitzer/IRS & 10.3$\micron$ & 2.81$\pm$0.08 & ... & ... & ... \\
Jan 14, 2009 & MMTAO/BLINC-MIRAC4 & 10.3$\micron$ & 2.63$\pm$0.10 & 1.49$\pm$0.06 & 1.14$\pm$0.04 & 1.30$\pm$0.01 \\
\hline
Nov 16, 1999 & Keck/OSCIR & N-band & 2.27$\pm$0.22 & 1.53$\pm$0.14 & 0.74$\pm$0.05 & 2.07$\pm$0.03 \\
Feb 27, 2004 & Spitzer/IRS & N-band & 2.35$\pm$0.07 & ... & ... & ... \\ 
Jan 14, 2009 & MMTAO/BLINC-MIRAC4 & N-band & 2.17$\pm$0.08 & 1.17$\pm$0.05 & 0.99$\pm$0.04 & 1.19$\pm$0.01 \\
\hline
Feb 27, 2004 & Spitzer/IRS & Si-5 & 2.62$\pm$0.08 & ... & ... & ... \\
Sep 27, 2004 & Gemini/Michelle & Si-5 & 3.26$\pm$0.16 & 1.95$\pm$0.10 & 1.31$\pm$0.07 & 1.49$\pm$0.03 \\
Jan 14, 2009 & MMTAO/BLINC-MIRAC4 & Si-5 & 2.42$\pm$0.09 & 1.32$\pm$0.05 & 1.10$\pm$0.04 & 1.22$\pm$0.02 \\

\enddata
\tablecomments{The Spitzer/IRS and MMTAO/BLINC-MIRAC4 data are spectra that have been converted to the photometric systems of 10.3$\micron$ (IRTF/MIRAC2), N-band (Keck/OSCIR) and Si-5 (Gemini/Michelle) as described in Section 3.2.  The Keck/OSCIR photometry of UY Aur is from \citet{McCabe}.  Their data is presented in magnitudes, which we convert to Jy using the zero-mag flux of 33.99 Jy for $\alpha$ Lyr presented in their paper.}

\label{Photometry}
\end{deluxetable}

\clearpage

\begin{deluxetable}{lcccccccccccc}
\tabletypesize{\scriptsize}
\tablecaption{Silicate Shape/Strength Measurements}
\tablewidth{0pt}
\tablehead{
\colhead{Spectrum} &
\colhead{$S_{9.9}$} &
\colhead{$S_{11.25}/S_{9.9}$}

}

\startdata
UY Aur (Spitzer unresolved) & 1.53$\pm$0.01 & 0.86$\pm$0.01 \\
UY Aur (MMTAO combined) & 1.36$\pm$0.05 & 0.83$\pm$0.04 \\
UY Aur A (MMTAO) & 1.59$\pm$0.07 & 0.75$\pm$0.04 \\
UY Aur B (MMTAO) & 1.14$\pm$0.06 & 0.92$\pm$0.06 \\

\enddata

\label{shape/strength}
\end{deluxetable}

\clearpage

\clearpage

\clearpage

\clearpage

\clearpage

\clearpage

\clearpage

\clearpage

\clearpage

\clearpage

\clearpage

\clearpage

\clearpage

\bibliographystyle{apj}
\bibliography{database}

\end{document}